\documentclass[a4paper,10pt]{article}   
\usepackage{osajnl2} 
\usepackage[]{graphicx,amsmath}
\begin{document}

\title{Two-dimensional point spread matrix of layered metal-dielectric imaging elements}
\author{Rafa{\l}~Koty{\'n}ski,$^{1,*}$ Tomasz Antosiewicz,$^{2}$ Karol~Kr{\'o}l,$^{1}$ and Krassimir Panajotov$^{3,4}$}
\address{$^{1}$Faculty of Physics, University of Warsaw,  Pasteura 7, 02-093 Warsaw, Poland}
\address{$^{2}$Interdisciplinary Centre for Mathematical and Computational Modelling, University of Warsaw, Pawinskiego 5A, 02-106 Warsaw, Poland}
\address{$^{3}$Department of Applied Physics and Photonics, Vrije Universiteit Brussel, Pleinlaan 2, 1050 Brussels, Belgium}
\address{$^{4}$Institute of Solid State Physics, 72 Tzarigradsko Chaussee Blvd. 1784, Sofia, Bulgaria}
\email{$^{*}$rafal.kotynski@fuw.edu.pl}

\begin{abstract}
We describe the change of the spatial distribution of the state of polarisation occurring during two-dimensional imaging through a multilayer and in particular through a layered metallic flat lens. Linear or circular polarisation of incident light is not preserved due to the difference in the amplitude transfer functions for the TM and TE polarisations. In effect, the transfer function and the point spread function that characterize 2D imaging through a multilayer both have a matrix form and cross-polarisation coupling is observed for spatially modulated beams with a linear or circular incident polarisation. The point spread function in a matrix form is used to characterise the resolution of the superlens for different polarisation states. We demonstrate how the 2D PSF may be used to design a simple diffractive nanoelement consisting of two radial slits. The structure assures the separation of non-diffracting radial beams originating from two slits in the mask and exhibits an interesting property of a backward power flow in between the two rings.
\end{abstract}

\maketitle



\section{INTRODUCTION}
\label{sect:intro}
In 2000, Pendry~\cite{pendry2000nrm} showed that a simple silver slab is capable of imaging in the near-field with resolution beyond the diffraction limit for the near-UV wavelength range. The operation of a similar but asymmetric\cite{anantharamakrishna2002aln} flat lens was later experimentally verified~\cite{melville2005sri,fang2005sdl}. In these works, the mechanism leading to  super-resolution is related to the appearance of a surface plasmon polariton (SPP) mode, which enables transfer of the evanescent part of the spatial spectrum through the slab. Other ways of achieving superlensing include the use of photonic crystals showing negative refraction~\cite{notomi2000tlp,zhang2006lst,luo2003sip,grbic2004odl,zhang2007can}. Both losses and the cut-off wavelength of the SPP mode limit the operation range to sub-wavelength distances only\cite{webb2004mnr,smith2002lsd}. This obstacle has been tackled to some degree in several ways. One of them is by breaking the silver slab into a metallic multilayer \cite{anantharamakrishna2002aln,wood2006} with effective anisotropic properties and strong coupling of the SPP modes between neighbouring layers. Otherwise the loss may be compensated by the use of media with gain \cite{anantharamakrishna2003raa,ponizovskaya2007mni}. Notably, according to the the effective medium model, a layered lens is approximated with a uniaxially anisotropic uniform slab~\cite{wood2006}. Uniaxial crystals are useful in the engineering of spatially varying polarisation states and optical vortices~\cite{Fadeyeva:oe2010,Wang:oe2010} and are a convenient medium for coupling cylindrical beams~\cite{Zhan:aop2009}.

Until the introduction of the concept of a perfect flat lens
with either a single layer~\cite{pendry2000nrm,Ramakrishna:JMO-49-1747,melville2005sri,fang2005sdl} or with multiple layers~\cite{anantharamakrishna2003raa}, it
was rather unusual to regard multilayers as spatial
imaging systems or linear spatial filters. Instead, the conventional
characteristics of a multilayer include the dependence of
transmission and reflection coefficients on the angle of incidence and polarisation, as well as the photonic band structure in case of periodic stacks. However, in order to describe
the resolution of an imaging system consisting of a multilayer in a systematic way, it is convenient to refer to the theory of linear shift invariant systems (LSI, also termed as linear isoplanatic systems~\cite{Saleh,GoodmanFourierOptics}) and such a description already gained  considerable interest~\cite{Moore:josaa2008,Scalora:OE09,Kotynski:ol2010,Kotynski:jopa2009,Kotynski:oer2010}. In this paper, metal-dielectric
multilayers (MDM) are regarded as LSI systems and a layered superlens is a one-dimensional spatial filter characterised by the point spread function (PSF). This approach
may facilitate the application of plasmonic elements to optical signal processing which is currently capturing increasing
research interest~\cite{Lee-Lalanne-Fainman:ao2010}. In the present paper we introduce a vectorial description of a linear spatial filter consisting of a multilayer, which enables accounting for polarisation coupling that occurs during imaging.
Therefore we examine features specific to 2D imaging when the description of the system does not trivially decouple into independent TE and TM polarisations.


\begin{figure}
\begin{center}
\includegraphics[width=10cm]{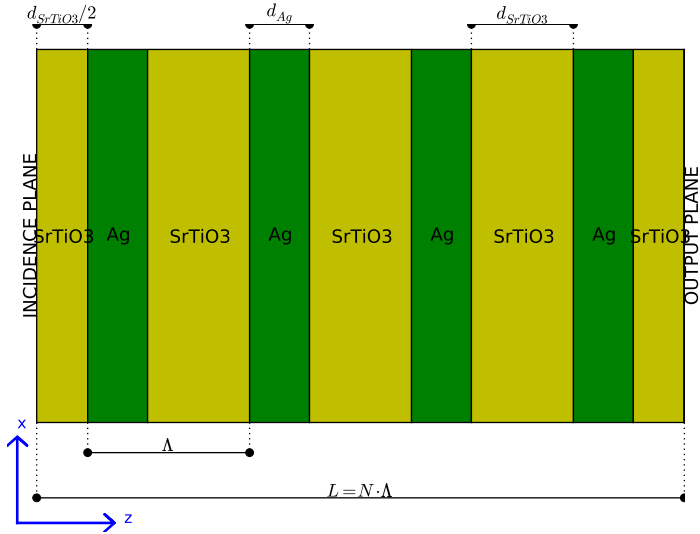}
\end{center}
\caption{Schematic of a layered metal-dielectric periodic lens with $N=4$ periods. It is further assumed that $\lambda =430\text  {nm}$, $\Lambda =57.5\text  {nm}$, $N=10$, $d_{\text{Ag}}=0.37\Lambda$, $d_{\text{SrTiO}_3}=0.63\Lambda$, $n_{\text{Ag}}=0.04+2.46i$, $n_{\text{SrTiO}_3}=2.674+0.027i$.}
\label{fig.lens-schem}
\end{figure}

\section{Background}
We analyse coherent imaging through a multilayer consisting of linear and isotropic materials. Further, we focus on metal-dielectric multilayers. A schematic of a periodic multilayered superlens is shown in Fig.~\ref{fig.lens-schem}. The slab boundaries are infinite and parallel to each another, and perpendicular to the $\hat z$ direction. Such an imaging element is an example of an LSI system which realises linear spatial filtering of the TE ($E_z=0$) or TM ($H_z=0$) -polarised incident wavefunction. Notably, diffraction in air and propagation through a metallic or dielectric slab are simple examples of linear spatial filters of the incident field, while a layered superlens is a more sophisticated one and may be seen as a superposition of such simpler systems, however it is still a linear spatial filter. From an engineering perspective, multiple degrees of freedom of a multilayer provide some leeway to perform point spread function optimisation of the response of the spatial filter.
The distribution of scalar electric or magnetic field components in the plane of incidence $\Psi(x,y)_{z=z_{inc}}$ and an output plane $\Psi(x,y)_{z=z_{out}}$, with a layered structure in between, are related through a convolution relation with the two-dimensional point spread function $\mathcal{H}(x,y)$,
\begin{equation}
\Psi(x,y)_{z=z_{out}}=(2\pi)^{-1} \mathcal{H}(x,y) \ast \Psi(x,y)_{z=z_{inc}}.
\end{equation}


Here $\ast$ denotes a two-dimensional convolution and we assume that propagation is in the $\hat z$ direction. The same relation transformed to the domain of spatial harmonics provides the definition of the transfer function (TF) which we denote as $\mathcal{\hat H}$,
\begin{equation}
\hat \Psi(k_x,k_y)_{z=z_{out}}= \mathcal{\hat H}(k_x,k_y) \cdot \hat \Psi(k_x,k_y)_{z=z_{inc}},
\end{equation}
where the hat represents the two-dimensional unitary Fourier transform, defined as
\begin{equation}
\begin{array}{c}\Psi(x,y)_z\end{array} =(2\pi)^{-1}\int\int \begin{array}{c}\hat \Psi(k_x,k_y)_z\end{array} e^{i( k_x x+k_y y) }dk_x dk_y.
\end{equation}
The PSF, related to the TF through the spatial Fourier transform, provides precise information on reshaping of the optical signal due to diffraction combined with multiple reflections and refractions, and is direct and convenient to interpret. Resolution of the system depends on the bandwidth of the transmitted spatial spectrum and super-resolution in the near field occurs when the transfer function of the system does not vanish for the evanescent part of the spatial spectrum ($k_x^2>1$).

Until now, we assumed that the system is scalar. This is only true when the field is purely TE or TM-polarised or when the transfer function is the same for the TE or TM polarisations which basically occurs only for simple propagation in air. There exist at least two important situations when the field symmetry in our system decouples Maxwell's equations into these two polarisation states. One takes place for in-plane imaging (e.g. in the $\hat x-\hat z$ plane) of 1D field distributions, when imaging of scalar fields $(H_y(x), E_x(x), E_z(x))$ and $(E_y(x), H_x(x), H_z(x))$ are independent and stand for the TM and TE polarisations, respectively. The other important situation takes place for a cylindrical beam with the angular wavenumber equal to $m=0$, when the radial $(E_r(r), H_\phi(r), E_z(r))$ and angular $(H_r(r), E_\phi(r), H_z(r))$ polarisations are also examples of pure TM and TE polarisations. In both situations, the symmetry of the system requires that the incident field is only one-dimensional with the dependence on either $x$ or $r$. More complex incident field distributions are neither TE or TM polarised and due to the different transfer function for the TE and TM polarisations, are subject to a change in the polarisation state during imaging through a multilayer.

\section{2D imaging properties of layered systems in 3D}
 At a plane $z=z_0$, a monochromatic field is characterised by two polarisation components. For instance, for the linear or circular polarisation  these are  $E_x(x,y)_{z=z_0}$ and $E_y(x,y)_{z=z_0}$, or $E_{\sigma_+}(x,y)_{z=z_0}$ and $E_{\sigma_-}(x,y)_{z=z_0}$, respectively.

Initially, let us assume knowledge of the one-dimensional transfer function (TF) for the TE and TM polarisations, denoting them  as $ \mathcal{\hat H}_{TE}(k_x)$ and  $\mathcal{\hat H}_{TM}(k_x)$, respectively. These TFs may be easily determined using the transfer matrix method from the complex transmission coefficients of propagating and evanescent planewaves with the wavevector components in the $x-z$ plane. We introduce denotations for the mean and difference of these two TFs, $\mathcal{\hat  H}_m=(\mathcal{\hat H}_{TM}+\mathcal{\hat H}_{TE})/2$, and $\mathcal{\hat  H}_\delta=(\mathcal{\hat H}_{TM}- \mathcal{\hat  H}_{TE})/2$, which are more convenient for the description of 2D imaging.

More generally, the incident and output fields are not scalar but vectorial and the 3D transmission of 2D field distributions through the multilayer transforms the spatial distribution of the polarisation state. Obviously in a planar geometry, the description of the system decouples into TE and TM polarisations in a simple way.
In 3D, the polarisation-dependent 2D transfer function $\mathcal{\hat  H}$ and point spread function $\mathcal{H}$, take the form of $2\times2$ matrices responsible for the transfer and coupling of both polarisations.
We will now derive these functions for linear and circular polarisations. In cylindrical coordinates, the following polarisation dependent relations can be written between the incident and output fields for linear polarisations,
\begin{equation}
\left[\begin{array}{c}\hat E_x(k_r,k_\varphi)\\ \hat E_y(k_r,k_\varphi)\end{array}\right]_{z=z_{out}} =\mathcal{\hat H}_{lin}(k_r,k_\varphi) \cdot \left[\begin{array}{c}\hat E_x(k_r,k_\varphi)\\ \hat E_y(k_r,k_\varphi)\end{array}\right]_{z=z_{in}},\label{eq.def_hlin}
\end{equation}
and for circular polarisations
\begin{equation}
\left[\begin{array}{c}\hat E_{\sigma_+}(k_r,k_\varphi)\\ \hat E_{\sigma_-}(k_r,k_\varphi)\end{array}\right]_{z=z_{out}} = \mathcal{\hat H}_{circ}(k_r,k_\varphi) \cdot \left[\begin{array}{c}\hat E_{\sigma_+}(k_r,k_\varphi)\\ \hat E_{\sigma_-}(k_r,k_\varphi)\end{array}\right]_{z=z_{in}},\label{eq.def_hcirc}
\end{equation}
respectively, where $ \mathcal{\hat H}_{lin}(k_r,k_\varphi)$ and $\mathcal{\hat  H}_{circ}(k_r,k_\varphi)$ are TFs in the matrix form which we need to determine. The right sides of Eqs.~(\ref{eq.def_hlin}) and~(\ref{eq.def_hcirc}) include matrix-vector multiplications. Notably, using the Jones calculus, it is straightforward to modify Eqs.~(\ref{eq.def_hlin}) and~(\ref{eq.def_hcirc}) and obtain expressions for the cross-coupling between linear and circular polarisations as well.
Now, we will express these 2D TFs by the TFs for the TE and TM polarisations, by rotating the coordinate system and casting the 3D situation to a planar TE or TM case,

\begin{eqnarray}
\mathcal{\hat  H}_{lin}(k_r,k_\varphi)
& = & \left[\begin{array}{cc} \mathcal{\hat H}_{xx}(k_r,k_\varphi) & \mathcal{\hat  H}_{yx}(k_r,k_\varphi) \\  \mathcal{\hat H}_{xy}(k_r,k_\varphi) & \mathcal{\hat  H}_{yy}(k_r,k_\varphi) \end{array} \right]\\
& = & R(k_\varphi)\cdot\left[\begin{array}{cc} \mathcal{\hat H}_{TM}(k_r) & 0 \\ 0 &  \mathcal{\hat H}_{TE}(k_r) \end{array} \right] \cdot R(-k_\varphi) \\
& = &  \mathcal{\hat H}_m(k_r)\cdot \left[\begin{array}{cc}1 & 0 \\ 0 & 1 \end{array} \right] + \mathcal{\hat H}_\delta(k_r)\cdot \left[\begin{array}{cc}cos(2 k_\varphi) & sin(2 k_\varphi)\\sin(2 k_\varphi) & -cos(2 k_\varphi) \end{array} \right],\label{eq.exp_hlin}
\end{eqnarray}
where $R$ is the rotation matrix and $\mathcal{\hat H}_{xx}(k_r,k_\varphi)$ may be understood as the transfer function linking the x-polarised component of the input signal to the x-polarised component of the output signal,  $\mathcal{\hat H}_{xy}(k_r,k_\varphi)$ as the transfer function linking the x-polarised input signal to the y-polarised component of the output signal, etc. The mean value of the TE and TM TFs $\mathcal{\hat H}_m(k_r)$ stands for the angularly independent term of the transfer function $\mathcal{\hat  H}_{lin}$. On the other hand, the difference between the TM and TE TFs, $ \mathcal{\hat H}_\delta(k_r)$  appears in the transfer function $\mathcal{\hat  H}_{lin}$ with an angular dependence and is responsible for polarisation coupling. This term vanishes when TE and TM polarisations become degenerated, for instance for diffraction in air. Conversely, plasmon-assisted transmission of evanescent waves is possible for TM polarisation only, and this term then results in strong polarisation coupling in 2D.

A similar expression for the transfer matrix for circular polarisations is obtained from Eq.~(\ref{eq.exp_hlin}) using the relations between linear and circular polarisations, $E_{\sigma_+}=(E_x+i E_y)/\sqrt{2}$ and $E_{\sigma_-}=(E_x-i E_y)/\sqrt{2}$,
\begin{eqnarray}
\mathcal{\hat H}_{circ}(k_r,k_\varphi)
& = &  \left[\begin{array}{cc}\mathcal{\hat H}_{\sigma_+\sigma_+}(k_r,k_\varphi) &  \mathcal{\hat H}_{\sigma_-\sigma_+}(k_r,k_\varphi) \\  \mathcal{\hat H}_{\sigma_+\sigma_-}(k_r,k_\varphi) &  \mathcal{\hat H}_{\sigma_-\sigma_-}(k_r,k_\varphi) \end{array} \right]\\
& = &  \mathcal{\hat  H}_m(k_r)\cdot \left[\begin{array}{cc}1 & 0 \\ 0 & 1 \end{array} \right] +\mathcal{\hat  H}_\delta(k_r)\cdot \left[\begin{array}{cc}0& exp(2 i k_\varphi)\\exp(-2 i k_\varphi) & 0\end{array} \right],
\end{eqnarray}

In order to derive the corresponding 2D point spread functions, we will use the following decomposition for the 2D Fourier transform of a function separable in the angular coordinate system. For $g(r,\varphi)=g_r(r)\cdot g_\varphi(\varphi)$, the corresponding Fourier transform ${\hat  g}(k_r,k_\varphi)$ is equal to~\cite{GoodmanFourierOptics},

\begin{equation}
{\hat  g}(k_r,k_\varphi)=\sum_{n=-\infty}^{+\infty}(-i)^n c_n \cdot exp(i n k_\varphi) \cdot {\mathcal{\mathsf{H}}}_n\left\lbrace g_{r}(r)\right\rbrace,
\end{equation}

where, the expansion coefficients $c_n$ are equal to

\begin{equation}
c_n=\int_0^{2\pi} g_\varphi(\varphi)\cdot exp(-i n \varphi)d\varphi,
\end{equation}

and $\mathsf{H}_n$ denotes the n-th order Hankel transform,

\begin{equation}
{\mathsf{H}}_n\left\lbrace g_{r}(r)\right\rbrace =\int_0^\infty  g_{r}(r)\cdot  J_n(k_r r)\cdot r \cdot dr.
\end{equation}

This decomposition leads to the following expressions for the polarisation-dependent 2D point spread function of the layered system,

\begin{eqnarray}
\mathcal{H}_{lin}(r,\varphi)
& = & \left[\begin{array}{cc} \mathcal{H}_{xx}(r,\varphi) & \mathcal{H}_{yx}(r,\varphi) \\  \mathcal{H}_{xy}(r,\varphi) &  \mathcal{H}_{yy}(r,\varphi) \end{array} \right]
\\
& = & \mathcal{ H}_m(r)\cdot
\left[
\begin{array}{rr}
1&0\\0&1
\end{array}
\right]
-\mathcal{ H}_\delta(r)\cdot
\left[
\begin{array}{cc}
cos(2\varphi) & sin(2\varphi) \\ sin(2\varphi) & -cos (2\varphi)
\end{array}
\right]\label{eq.ir_xy}
\end{eqnarray}

\begin{eqnarray}
\mathcal{H}_{circ}(r,\varphi)
& = &  \left[\begin{array}{cc} \mathcal{H}_{\sigma_+\sigma_+}(r,\varphi) &  \mathcal{H}_{\sigma_-\sigma_+}(r,\varphi) \\ \mathcal{H}_{\sigma_+\sigma_-}(r,\varphi) &  \mathcal{H}_{\sigma_-\sigma_-}(r,\varphi) \end{array} \right]\\
& = &
\mathcal{ H}_m(r)\cdot
\left[
\begin{array}{rr}
1&0\\0&1
\end{array}
\right]
-\mathcal{ H}_\delta(r)\cdot
\left[
\begin{array}{cc}
0 & exp( 2 i \varphi) \\ exp (-2 i \varphi) & 0
\end{array}
\right]\label{eq.ir_sgm}
\end{eqnarray}
where $\mathcal{ H}_m(r)={\mathsf{H}}_0\left\lbrace  \mathcal{\hat H}_m(k_r)\right\rbrace$ and $\mathcal{ H}_\delta(r)={\mathsf{H}}_2\left\lbrace  \mathcal{\hat H}_\delta(k_r)\right\rbrace$.

Summarising, the numerical calculation of the point spread functions in 2D requires the following steps: 1.~finding the 1D TE and TM transfer functions for a planar geometry using the transfer matrix method, 2.~determination of the 1D Hankel transforms of the 0-th and 2-nd orders of the mean and difference of these two TFs, respectively. 3.~Reconstruction of the point spread functions (\ref{eq.ir_xy}) and (\ref{eq.ir_sgm}) including the angular dependence. Therefore, the majority of the numerical calculations is done in 1D. While we evaluate directly the Hankel transforms, it should be noted that there exist fast numerical algorithms which may be applied instead~\cite{Norfolk:oe2010}. The 0-th order Hankel transform links the rotationally invariant part of PSF to the respective rotationally invariant part of TF, which in turn is equal to the average of TE and TM one-dimensional TFs. On the other hand, the difference of the TM and TE one-dimensional TFs is  responsible for the angularly dependent polarisation coupling and is related to the respective part of the point spread function through the second order Hankel transform.

\section{Polarisation coupling for linearly or circularly polarised incident beams}
In this section we calculate the TF-matrix and PSF-matrix for two-dimensional vectorial spatial filtering realized with a layered lens. We assume that the lens consists of silver and strontium titanate with a filling fraction of silver equal to $0.37$ and operates at a wavelength of $\lambda=430\text{nm}$. The structure is presented in Fig.~\ref{fig.lens-schem}. It is a low-loss  self-guiding superlens with a sub-wavelength FWHM of PSF for the TM polarisation in the order of $\lambda/10$ and it has been already thoroughly investigated in terms of transmission efficiency as well as of resolution for in-plane imaging~\cite{Kotynski:ol2010,Kotynski:oer2010}. The elementary cell consists of an Ag layer symmetrically coated with SrTiO$_3$. Strontium titanate is an isotropic material with a high refractive index $n=2.674+0.027i$ at  $\lambda=430\text{nm}$\cite{Palik}. The refractive index of silver at the same wavelength is equal to $n_{\text{Ag}}=0.04+2.46i$\cite{JohnsonChristy}. Further, we assume that the multilayer consists of $N=10$ periods with a thickness of $\Lambda=57.5\text{nm}$ each.

\begin{figure}
a)\includegraphics[height=6.5cm]{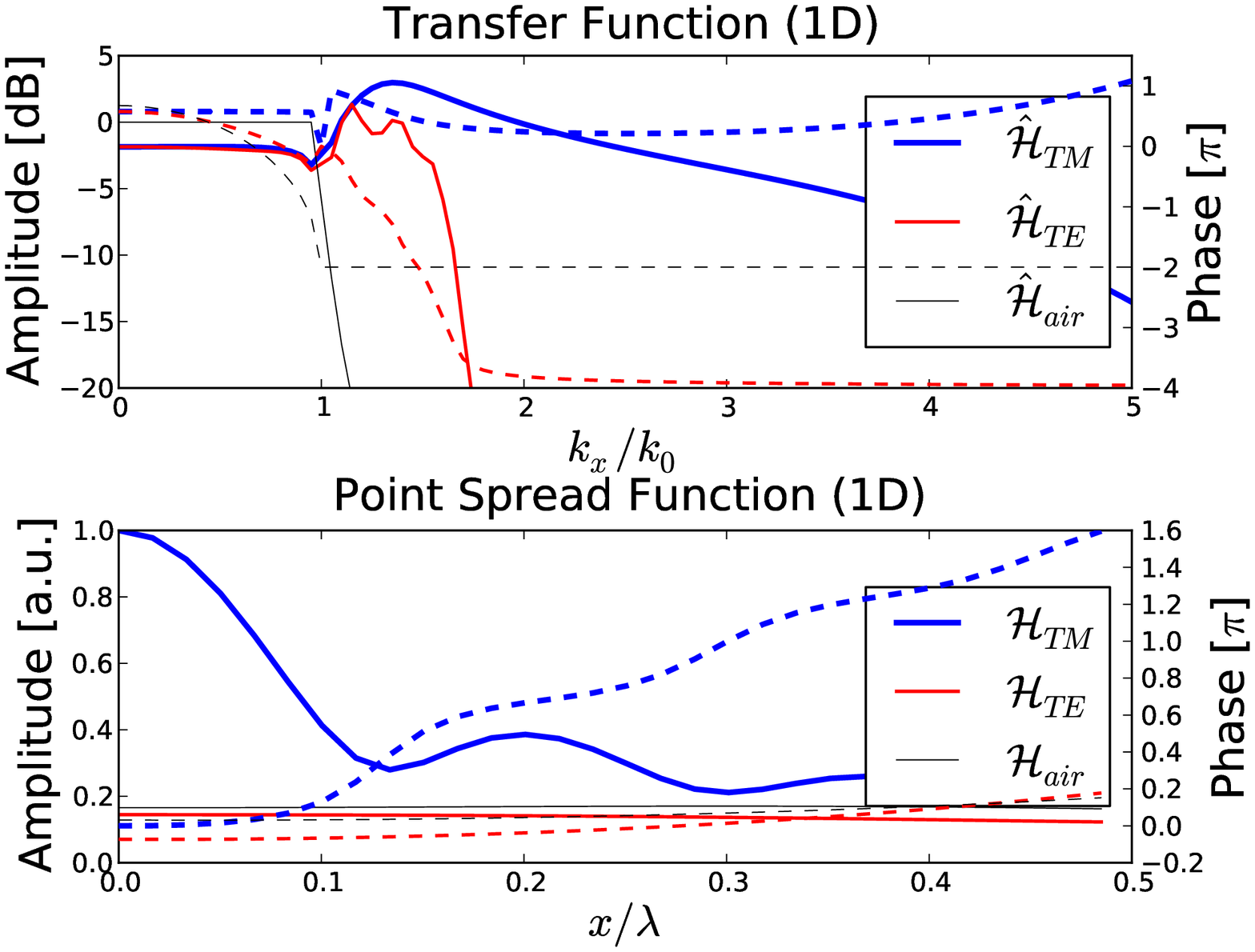}b)\includegraphics[height=6.5cm]{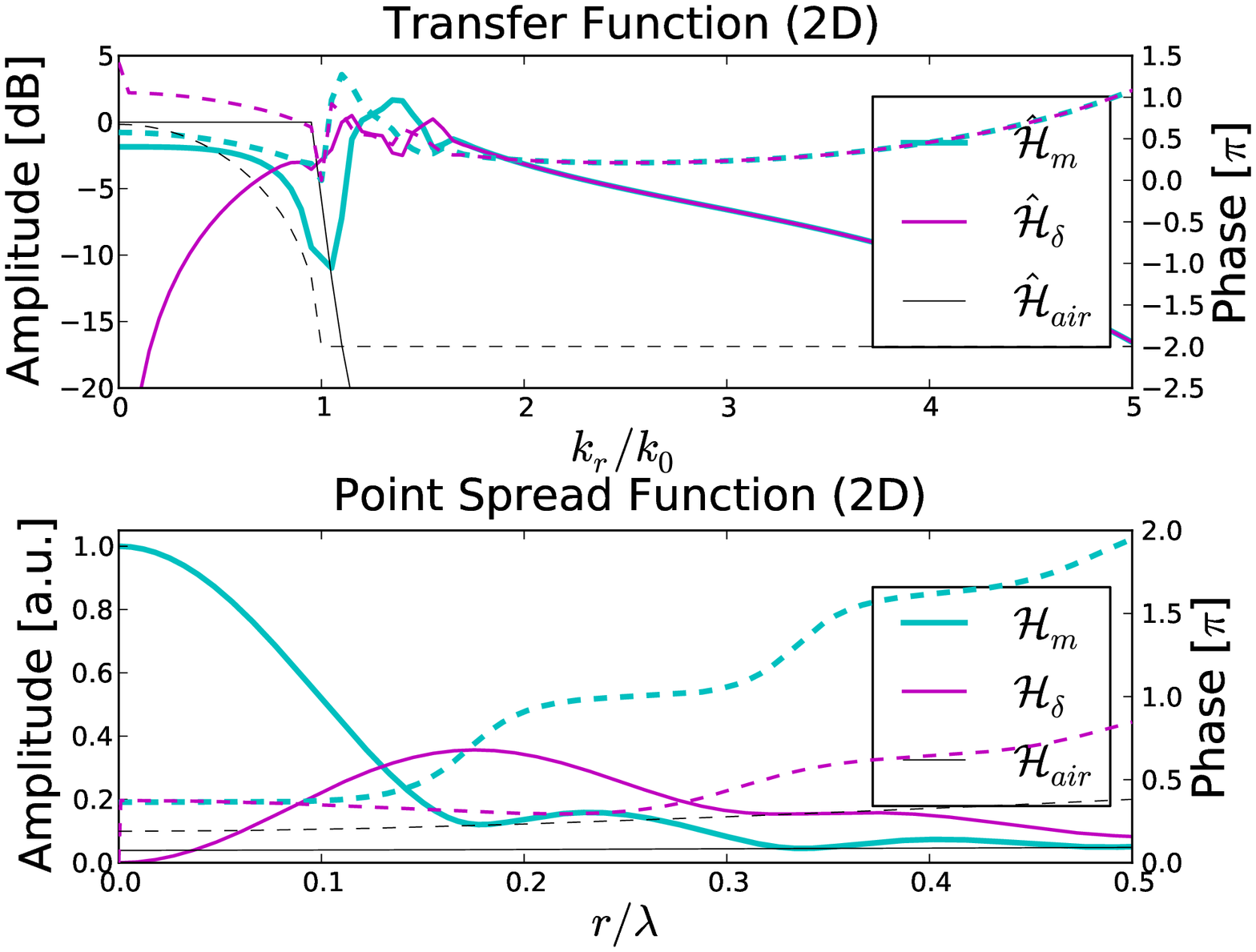}
\caption{a)~One-dimensional TF and PSF of the multilayer for in-plane propagation: $\mathcal{\hat H}_{TM}$,$\mathcal{\hat H}_{TE}$~-- TF for the TM and TE polarisations; b)~Radial cross-section of two-dimensional TF and PSFs of the multilayer for 3D propagation $\mathcal{\hat  H}_m=( \mathcal{\hat H}_{TM}+ \mathcal{\hat H}_{TE})/2$, $ \mathcal{\hat H}_\delta=(\mathcal{\hat  H}_{TM}-\mathcal{\hat H}_{TE})/2$. 1D and 2D PSFs and TF corresponding to  propagation in air for the same distance are also shown for comparison. Solid lines and dashed lines represent the amplitudes and phases of the respective complex functions.}
\label{fig.1dtf}
\end{figure}

In Fig.~\ref{fig.1dtf} in the top row we show the phase (dashed lines) and amplitude (solid lines) of the TFs: $\mathcal{\hat H}_{TM}(x)$,$\mathcal{\hat H}_{TE}(x)$, $\mathcal{\hat  H}_m(k_r)$,  $\mathcal{\hat H}_\delta(r)$, and the TF for propagation in air at the same distance. In the bottom row, the corresponding 1D PSF and the radial part of the 2D PSF are also presented. Our example illustrates some of the general properties of the TF - for normal incidence ($k_x=0$), we have $\mathcal{\hat H}_{TM}(0)=\mathcal{\hat H}_{TE}(0)=\mathcal{\hat  H}_m(0)$, and $\mathcal{\hat H}_\delta(r)=0$. On the other hand, for evanescent harmonics SPPs are only supported for the TM polarisation and therefore for a sufficiently large $k_x$ we have $|\mathcal{\hat H}_{TM}(k_x)|>>|\mathcal{\hat H}_{TE}(k_x)|$, and in effect $2\mathcal{\hat H}_{TM}\approx\mathcal{\hat  H}_m\approx \mathcal{\hat H}_\delta$. Therefore, the bandwidth and asymptotic behaviour of $\mathcal{\hat  H}_m$ and $\mathcal{\hat H}_\delta$ are the same as of $\mathcal{\hat H}_{TM}$~(see the solid line in Fig.~\ref{fig.1dtf}b, top for $k_r/k_0>1.8$). If the bandwidth of $\mathcal{\hat H}_{TM}$ extends to $k_x>>1$ and its phase is approximately constant, which are the necessary conditions for super-resolution, the same properties are also satisfied by $\mathcal{\hat  H}_m$ and $\mathcal{\hat H}_\delta$. Therefore, we expect to see super-resolution in 2D, even though 2D-imaging depends on both TE and TM polarisations and the TE polarisation itself does not allow for super-resolution. This observation is confirmed by the shape of 2D PSF presented in the same figure. We further see (the violet line in Fig.~\ref{fig.1dtf}b, bottom) that the polarisation coupling determined by $\mathcal{H}_\delta$ is zero at $r=0$ and smoothly increases with $k_r$ reaching a maximum at $r/\lambda\sim 0.18$ i.e. at (approximately) the first minimum of $\mathcal{  H}_m$.

\begin{figure}
a)\includegraphics[height=6.5cm]{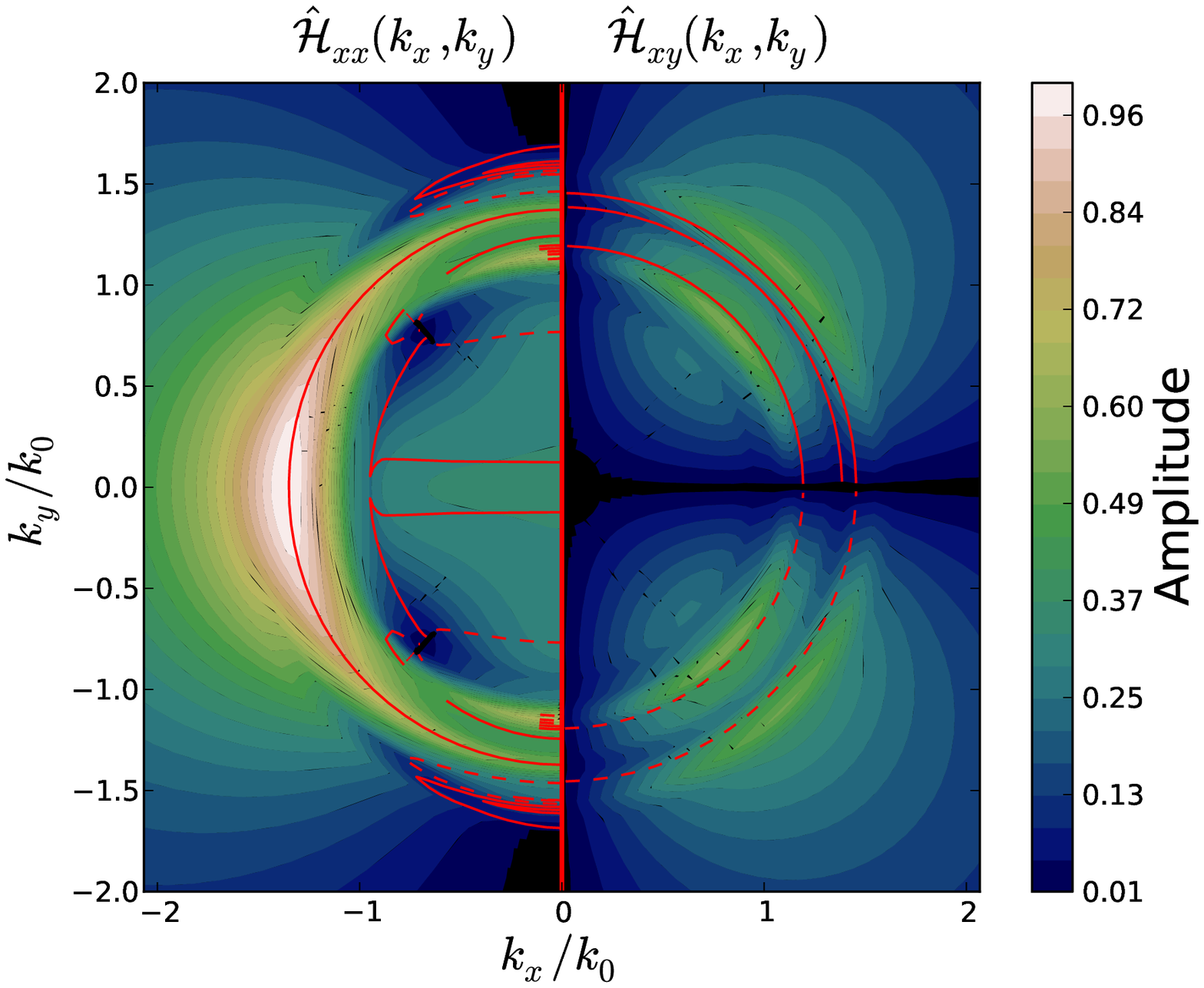}b)\includegraphics[height=6.5cm]{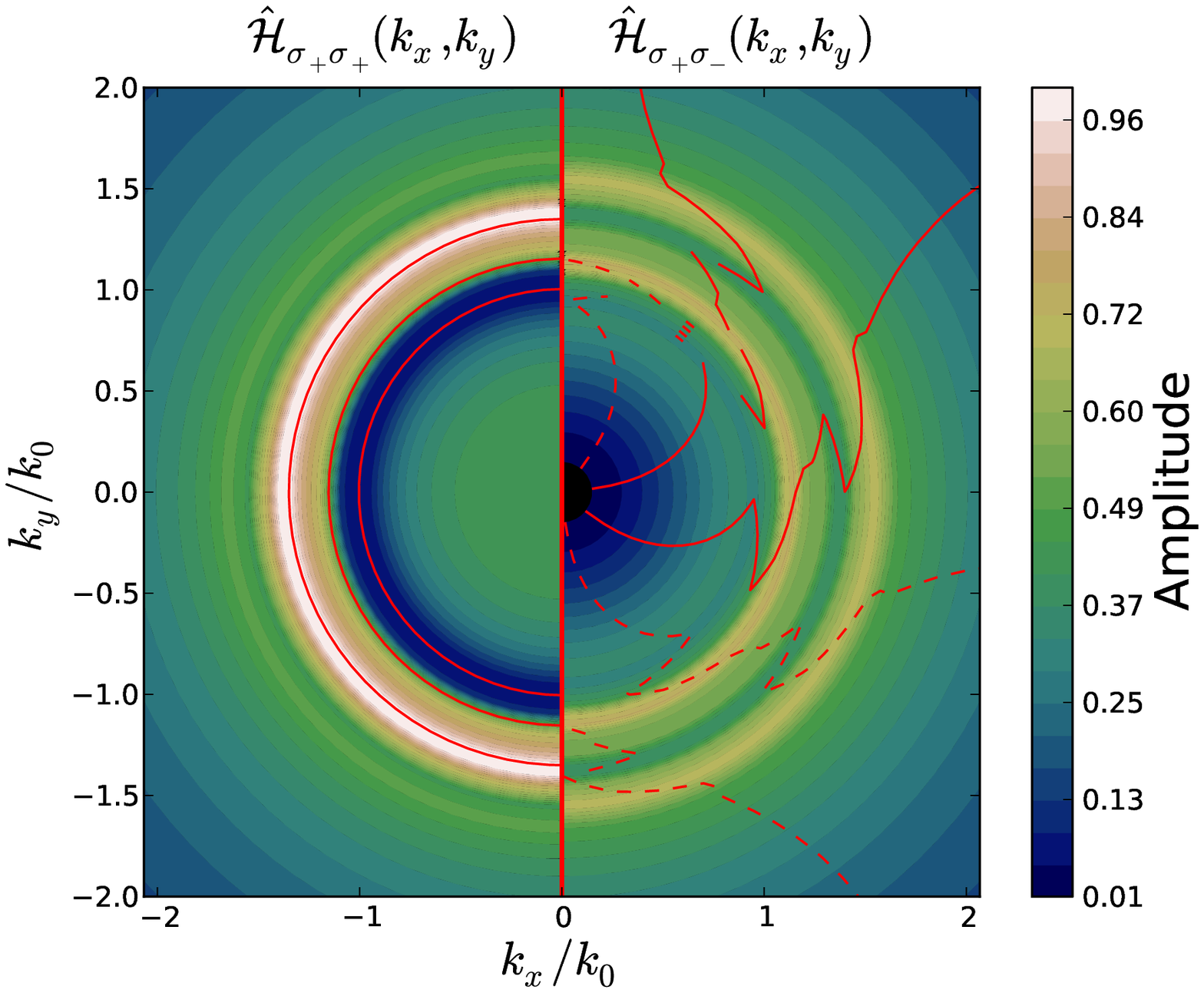}
\caption{ Two-dimensional TF-matrix of the multilayer. {a)}~Matrix elements ${\mathcal{\hat H}}_{xx}(k_x,k_y)$ and ${\mathcal{\hat H}}_{xy}(k_x,k_y)$ for linearly polarised light; {b)}~Matrix elements ${\mathcal{\hat H}}_{\sigma_+\sigma_+}(k_x,k_y)$ and ${\mathcal{\hat H}}_{\sigma_+\sigma_-}(k_x,k_y)$ for circularly polarised light. Phase isolines are drawn at the distances of $\pi/2$ and are changed between solid and dashed lines every $\pi$.}
\label{fig.2dtf}
\end{figure}

\begin{figure}
a)\includegraphics[height=6.5cm]{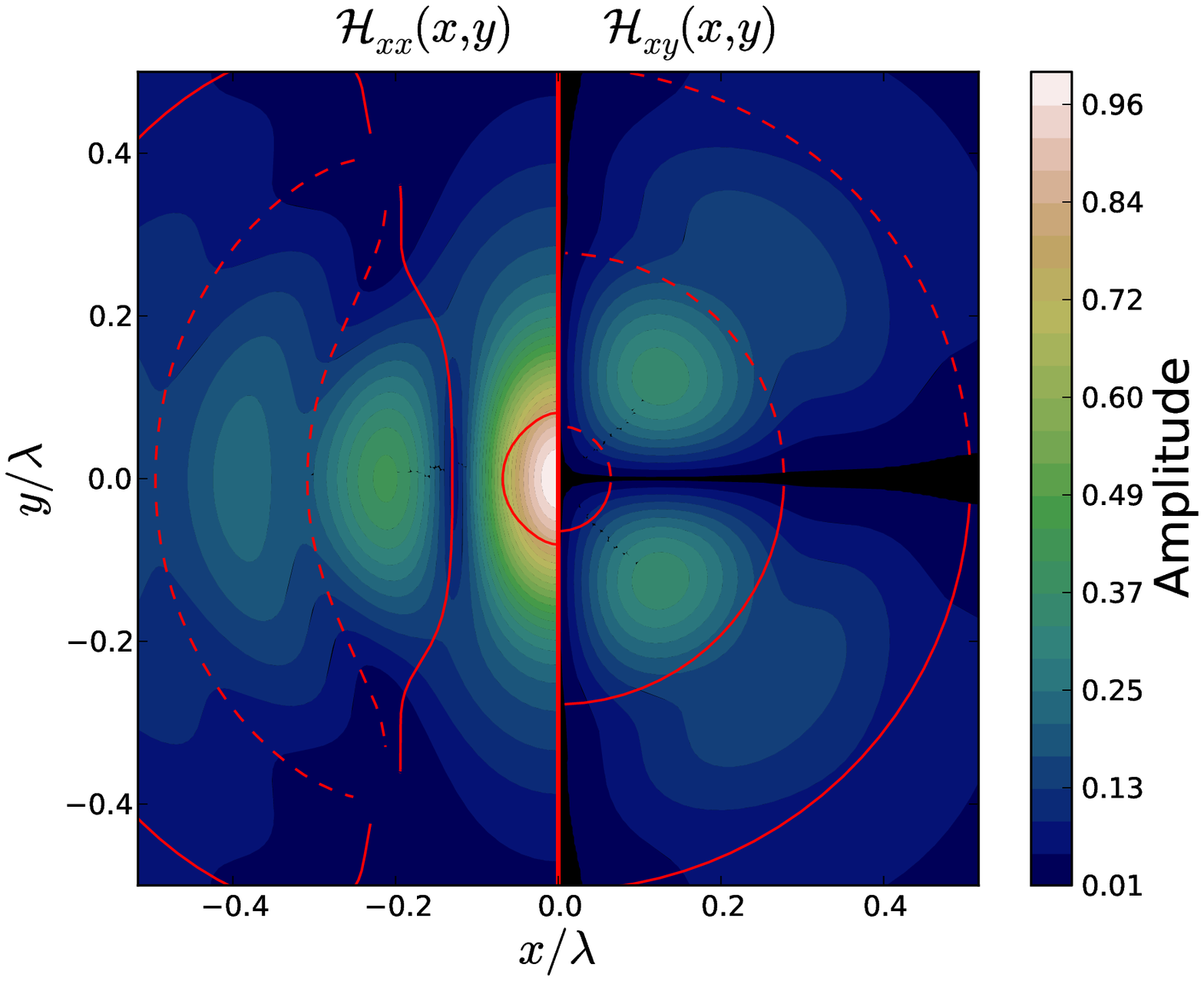}b)\includegraphics[height=6.5cm]{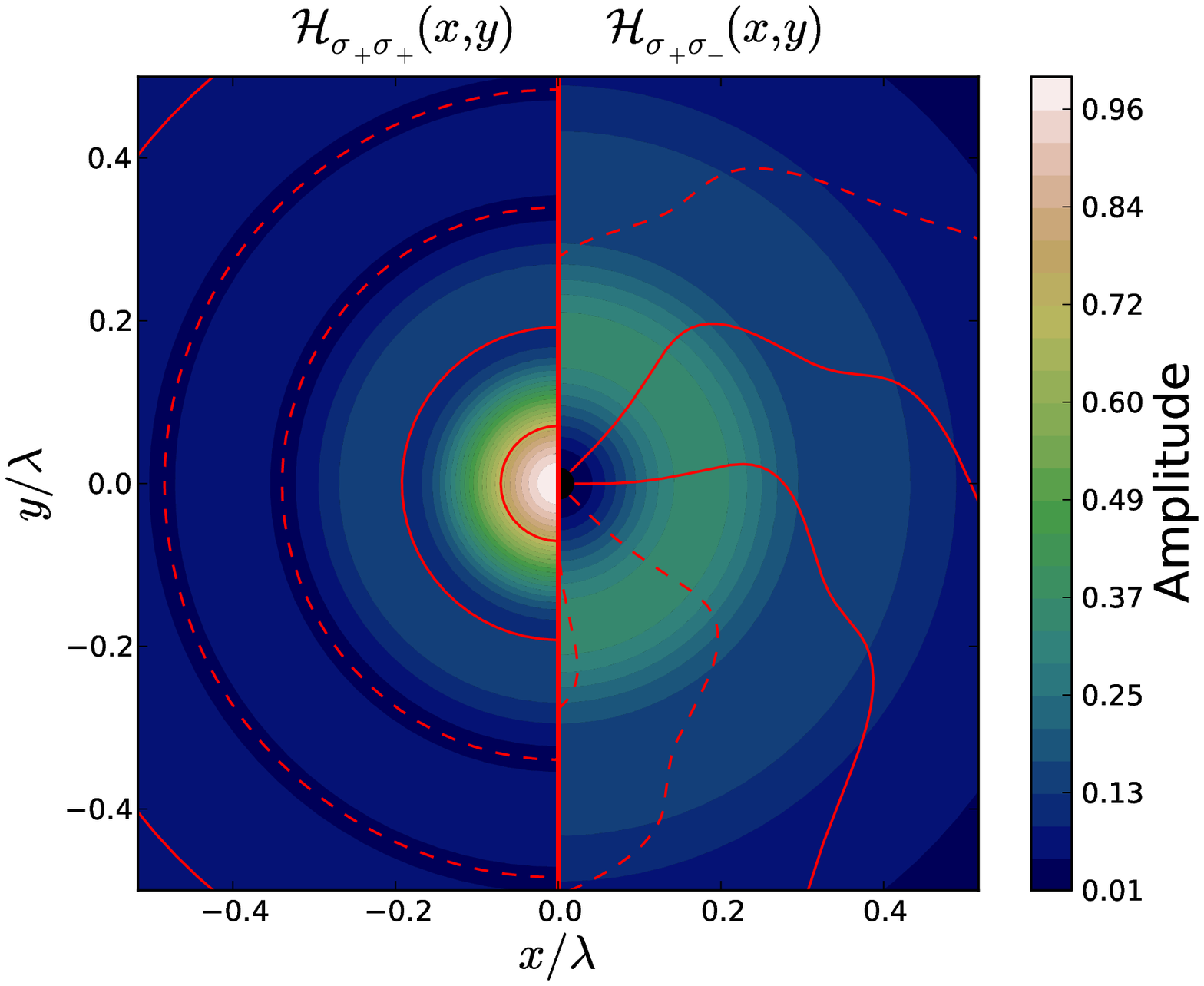}
\caption{ Two-dimensional PSF-matrix of the multilayer. {a)}~Matrix elements ${\mathcal{H}}_{xx}(x,y)$ and ${\mathcal{H}}_{xy}(x,y)$ for linearly polarised light;  {b)}~Matrix elements ${\mathcal{H}}_{\sigma_+\sigma_+}(x,y)$ and ${\mathcal{\hat H}}_{\sigma_+\sigma_-}(x,y)$ for circularly polarised light; Phase isolines are drawn at the distances of $\pi/2$ and are changed between solid and dashed lines every $\pi$.}
\label{fig.2dpsf}
\end{figure}

\begin{figure}
a)\includegraphics[height=6.5cm]{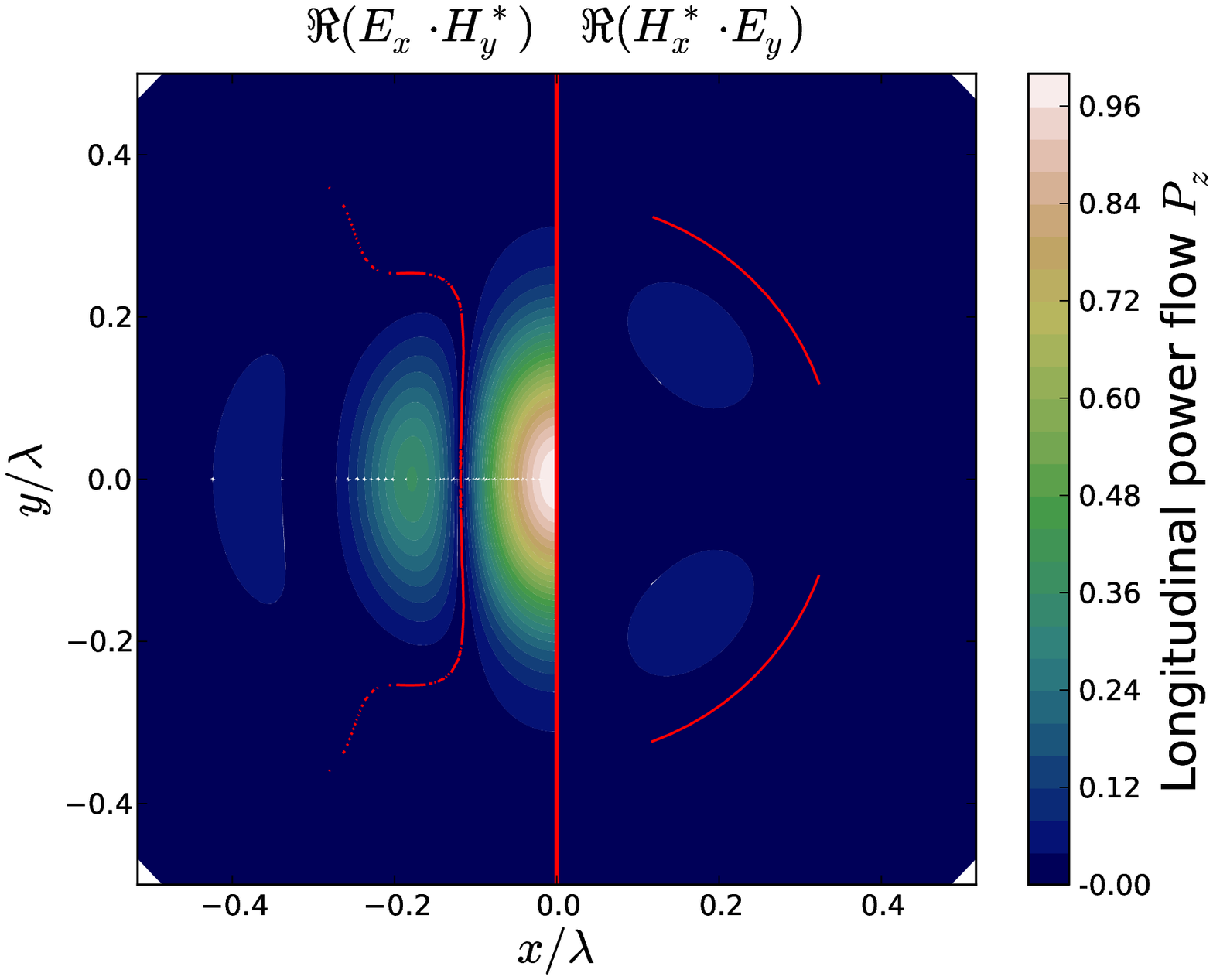}b)\includegraphics[height=6.5cm]{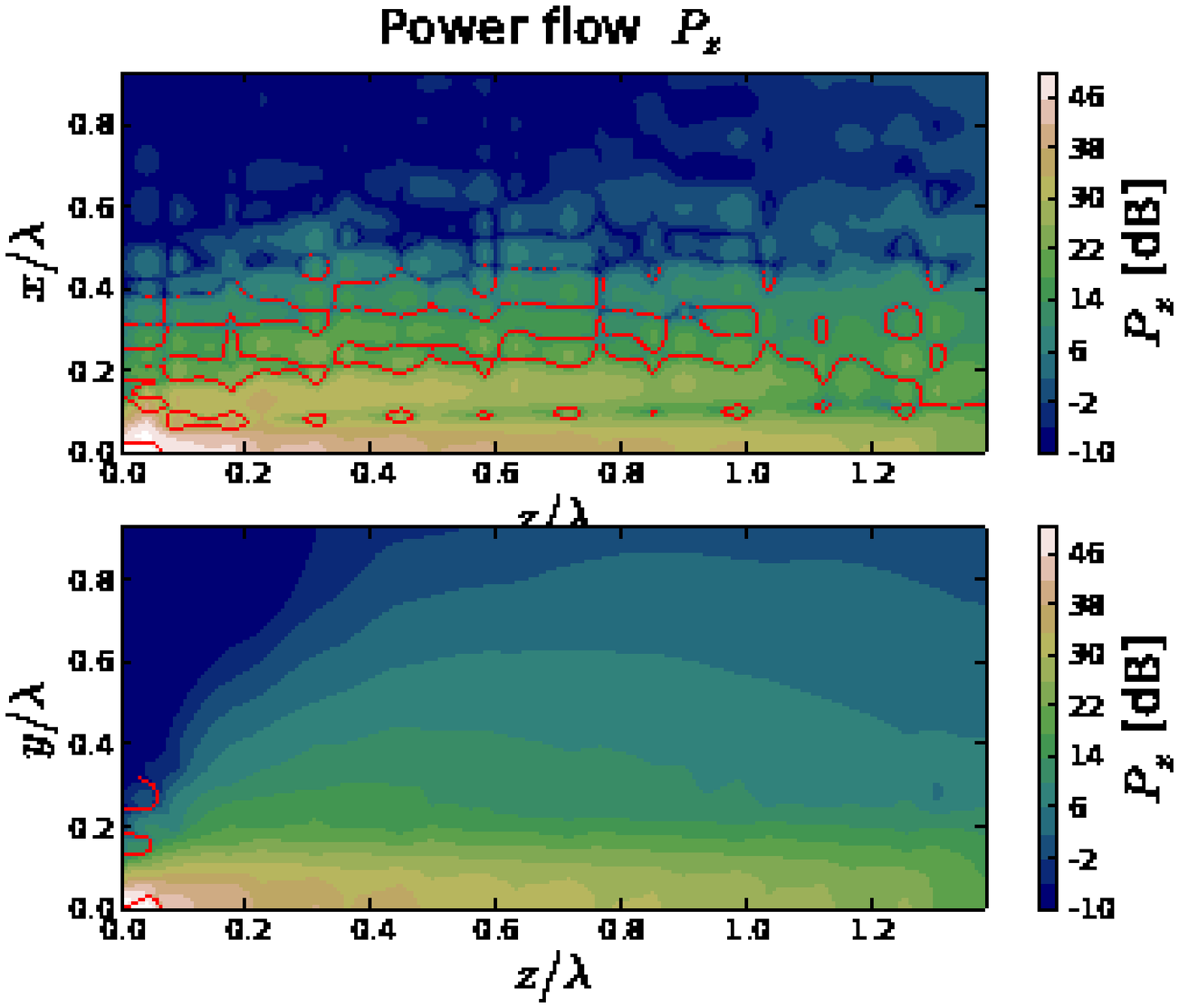}
\caption{{a)}~Contributions to the axial Poynting vector $P_z$ from both polarisations at the output plane, resulting from a linearly $x$-polarised incident point source; {b)}~Axial component of the Poynting vector $P_z$ in the $x-z$ and $y-z$ cross-sections inside the structure normalised with respect to $P_z(x=0,y=0,z=L)$. The simulations were performed using FDTD. The (red) lines separate areas with positive and negative direction of the power flow.}
\label{fig.2dfdtd}
\end{figure}

\begin{figure}
a)\includegraphics[height=6.5cm]{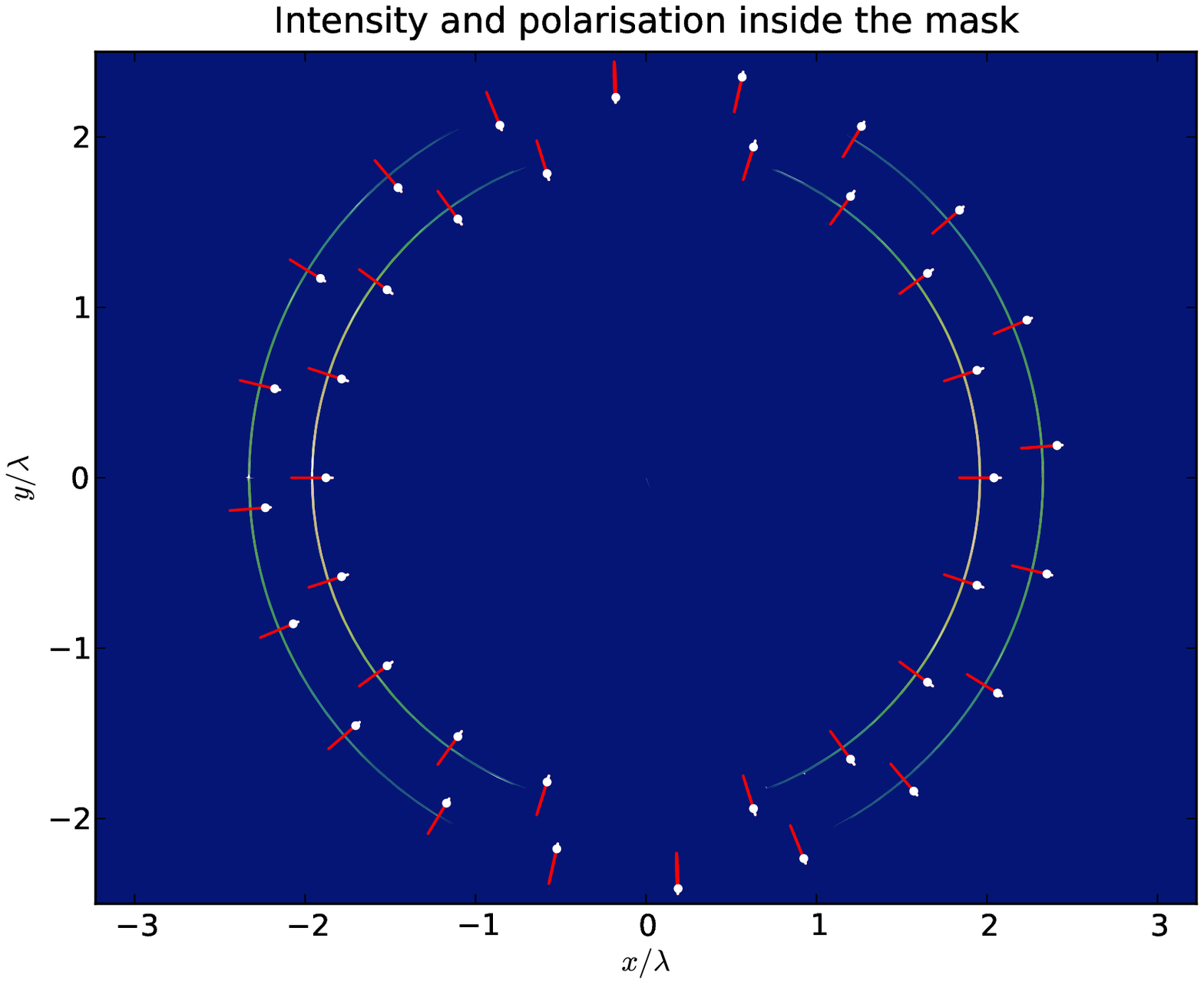}
b)\includegraphics[height=6.5cm]{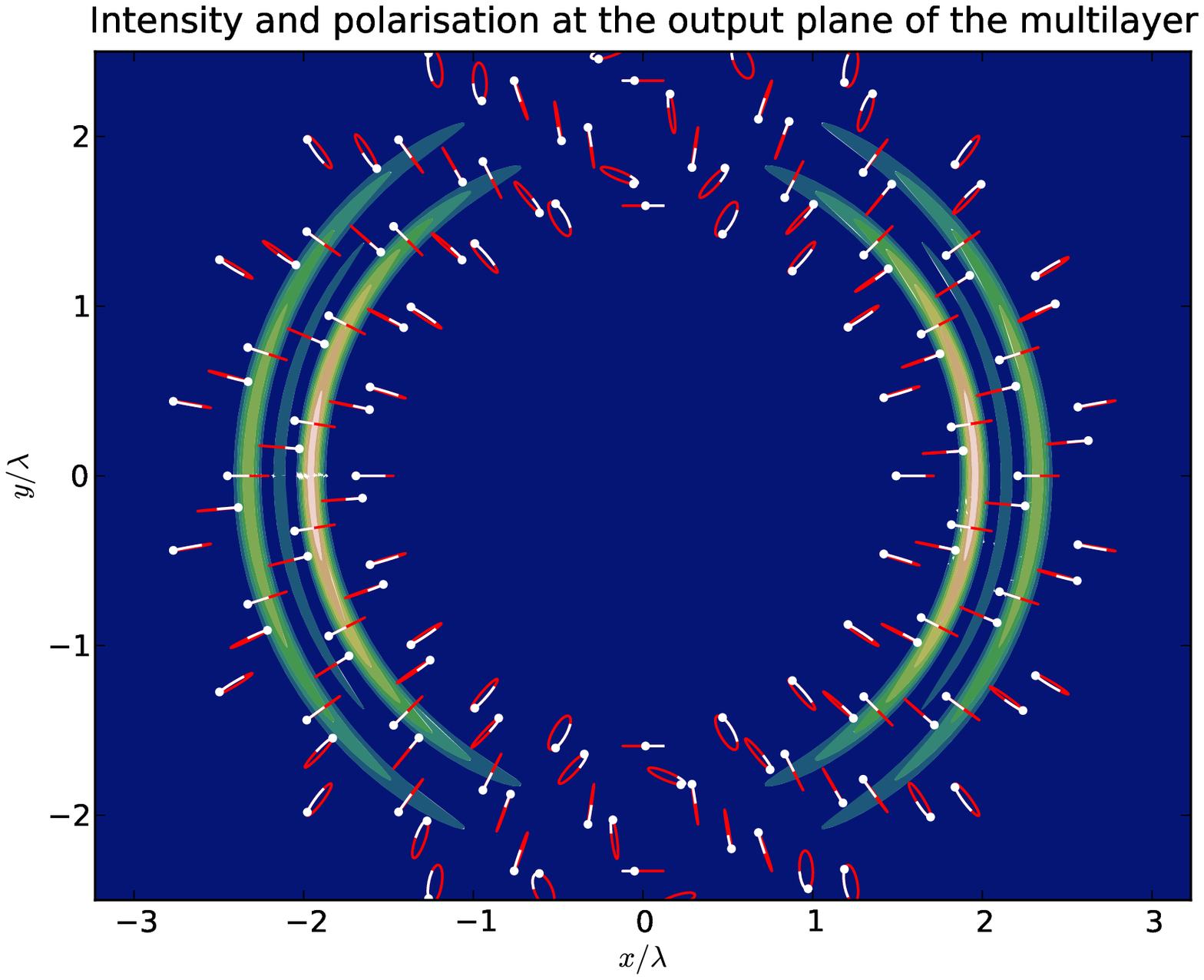}
\caption{Intensity of the electric field and the state of polarisation inside the mask ({a}) and at the output plane of the multilayer ({b}). The mask consists of a perfect conductor with two circular slits and is illuminated with a linearly x-polarised Gaussian beam. FDTD simulation, linear intensity mapping.}
\label{fig.twoslit_polarisation}
\end{figure}

\begin{figure}
\begin{center}
\includegraphics[height=6.5cm]{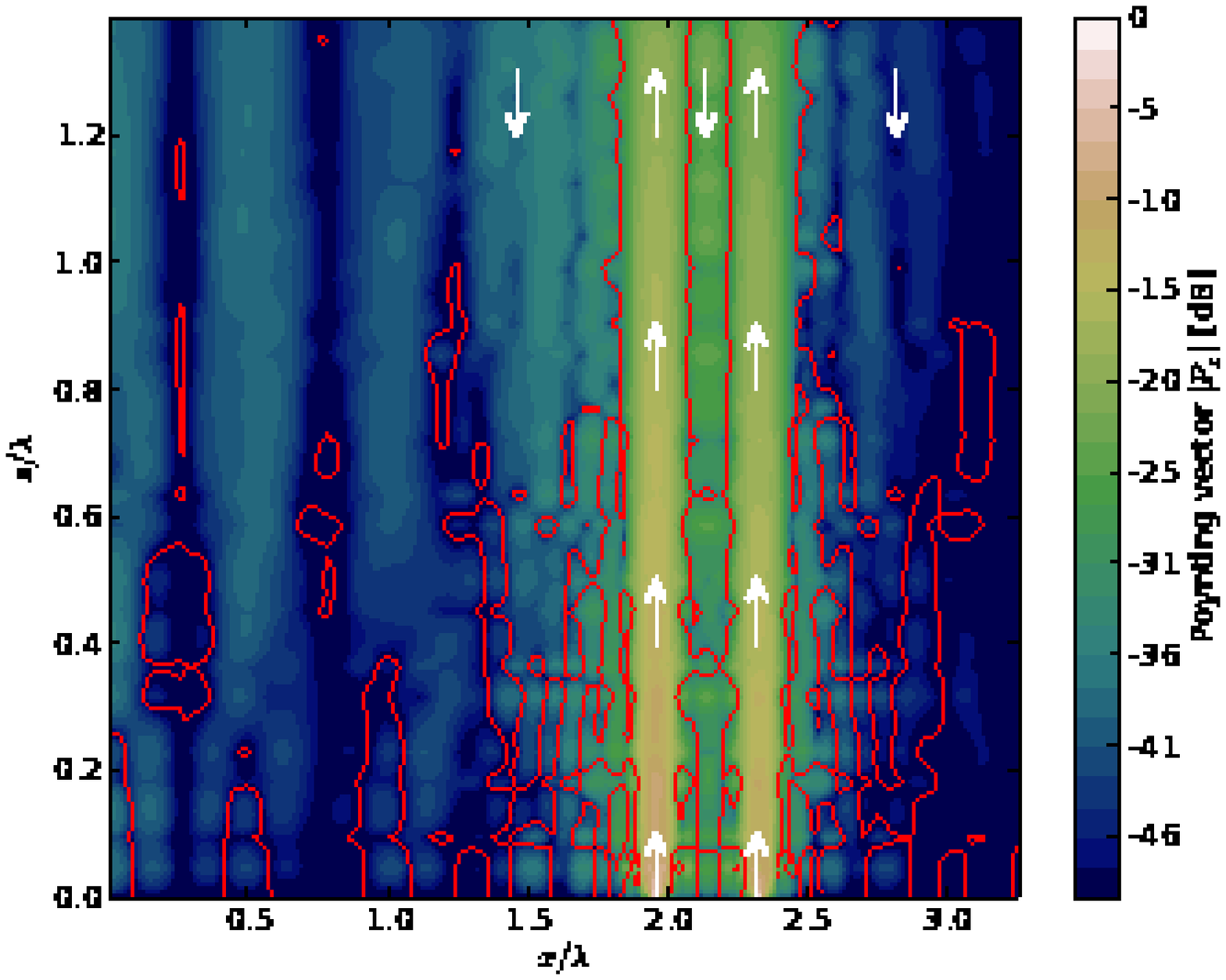}
\caption{Time-averaged Poynting vector component $\langle P_z \rangle$ in the $x-z$ cross-section of the structure (where $x=0$). The (red) lines separate areas with the positive and negative values of $P_z$ and the (white) arrows show the orientation of the energy flow in the vertical direction.}
\label{fig.twoslit_xz}
\end{center}
\end{figure}

In Figs.~\ref{fig.2dtf} and~\ref{fig.2dpsf} we show the same 2D TF and PSF as in Fig.~\ref{fig.1dtf}b, plotted however in the 2D space, including their angular dependence according to Eqns.~(\ref{eq.ir_xy}) and~(\ref{eq.ir_sgm}).  The polarisation-preserving and polarisation-coupled elements of the TF and PSF matrix are drawn for both linear and circular polarisations. Notably, the size of the PSF is sub-wavelength and the polarisation state is preserved in the central part of the PSF because the second order Hankel transform vanishes at the origin and the strongest coupling to the orthogonal polarisation is observed at some distance from the center. For circular polarisation, the term ${\mathcal{H}}_{\sigma_+\sigma_+}(x,y)$ depends on $\mathcal{ H}_m(r)$ and is rotationally symmetric while the term ${\mathcal{H}}_{\sigma_+\sigma_-}(x,y)$ depends on $\mathcal{ H}_\delta(r)$ and has a simple angular phase dependence (see Eq.\ref{eq.ir_sgm}, and Fig.~\ref{fig.2dpsf}b). For linear polarisation, the term ${\mathcal{H}}_{xx}(x,y)$  depends on both $\mathcal{ H}_m(r)$ and $\mathcal{H}_\delta(r)$ (see Eq.\ref{eq.ir_xy}, and Fig.~\ref{fig.2dpsf}a) and is not rotationally symmetric.

We have also calculated the same elements of the PSF-matrix using body-of-revolution FDTD and obtained good agreement with the previous results  calculated using the transfer matrix method and Hankel transforms. For comparison, the FDTD simulations are shown in Fig.~\ref{fig.2dfdtd} where we plot the axial energy flux resulting from a linearly x-polarised incident point source. Figure~\ref{fig.2dfdtd}a illustrates the contributions from the two linear polarisations to the Poynting vector $P_z$  at the output plane. These contributions may be directly matched to the 2D PSF presented in Fig.~\ref{fig.2dpsf}a. Figure~\ref{fig.2dfdtd}b shows  the Poynting vector $P_z$ in the $x-z$ and $y-z$ cross-sections of the structure and shows that the direction of polarisation has a major influence on the diffraction rate in orthogonal directions inside the structure.  The spatial distribution of  $P_z$ shown in Fig.~\ref{fig.2dfdtd}a with contributions from both polarisations is in good agreement with Fig.~\ref{fig.2dpsf}a.


Further, we demonstrate how the 2D PSF may be used to design a simple diffractive nanoelement. Using predicted properties of the 2D PSF we design a  pattern that could be imprinted into a metallic mask attached to the layered superlens. The pattern consists of two narrow radial slits and we use the PSF to determine the distance between the slits so that the interference between the polarisation-coupled elements of two images of slits is enhanced. From the 2D PSF shown in Fig.~\ref{fig.1dtf}b we find that for the radial separation of $r=0.18\lambda$ the maximum of the polarisation cross-coupling term $\mathcal{ H}_m(r)$ coincides with a minimum of the direct coupling term $\mathcal{ H}_\delta(r)$.  The FDTD simulation showing the operation of the proposed element is presented in Figs.~\ref{fig.twoslit_polarisation} and~\ref{fig.twoslit_xz}. A linearly polarised Gaussian beam is diffracted on the two circular slits in the mask. For the purpose of simplicity the mask is made of a perfect conductor (PEC) and the width of slits is equal to $4\text{ nm}$ which corresponds to $8$ grid-points of the FDTD 2D mesh. Inside the slits the wave becomes polarised perpendicularly to the slits. This is shown in Fig.~\ref{fig.twoslit_polarisation}a. It should be noted that inside the slits 
 the field is neither TE or TM polarised and $E_z\neq 0, H_z\neq 0$. 
The polarisation at the output plane is shown in Fig.~\ref{fig.twoslit_polarisation}b. In between the images of two slits there is a local maximum of intensity resulting from the interference of the 2D PSFs with the cross-polarised term eliminated. At the external side of the images of the slits, without interference, the cross-polarised terms of PSF are still present and the polarisation state varies considerably with respect to that within the slits. At the same time, in between the slits, the phase shift of $\mathcal{ H}_m(r)$ reaches approximately $\pi$ with respect to its central value, resulting in the reversal of the direction of power flow $P_z$. This reversal is demonstrated in Fig.~\ref{fig.twoslit_xz} which shows the cross-section of the multilayer along the $x-z$ axes. An inverted direction of $P_z$ may be seen along most of the propagation distance within the multilayer, although within the mask itself the direction of the power is clearly defined in the forward direction only.

\section{Conclusion}
Two-dimensional beams generally do not preserve their initial spatial distribution of the state of polarisation after passing through a layered flat lens. This is the direct result of different transfer functions corresponding to the TM and TE components of the spatial spectrum and the presence of both TM and TE components in the spatial spectrum of a 2D beam. We use the framework of the theory of linear shift-invariant systems to describe the imaging system and express the PSF $2\times2$-matrix for imaging of vectorial 2D wavefronts using the 0-th and 2-nd order Hankel transforms of planar transfer functions for the TE and TM polarisations.

We calculate the polarisation-preserving and cross-polarisation 2D PSF matrix elements of a layered superlens showing that the resolution in 2D is of the same order as the resolution for in-plane imaging with the TM polarisation. This comes in spite of the fact that both the TE and TM polarised spatial harmonics take part in the 2D imaging. Moreover, cross-polarisation coupling is observed for spatially modulated beams with a linear or circular incident polarisation.

We demonstrate how the 2D PSF may be used to design a simple diffractive nanoelement. The spatial separation of the PSF matrix elements for polarisation-preserving and polarisation-coupled imaging is used to design a mask that forms a wavefront which during propagation exhibits  destructive interference in the cross-polarisation term, assures the separation of non-diffracting radial beams originating from two slits in the mask and exhibits an interesting property of a backward power flow in between the two rings. This opens a possibility for further PSF engineering where the polarisation effects are accounted for.

\section*{Acknowledgments}
We acknowledge support from the Polish  research projects MNiI~N~N202~033237, and NCBiR~N~R15~0018~06 and the framework of COST MP0702 action.


\begin{thebibliography}{10}
\newcommand{\enquote}[1]{``#1''}

\bibitem{pendry2000nrm}
J.~B. Pendry, \enquote{Negative refraction makes a perfect lens,} Phys. Rev.
  Lett. \textbf{85}, 3966--3969 (2000).

\bibitem{anantharamakrishna2002aln}
S.~A. Ramakrishna, J.~B. Pendry, D.~Schurig, and D.~R. Smith, \enquote{The
  asymmetric lossy near-perfect lens,} J. Mod. Opt. \textbf{49}, 1747--1762
  (2002).

\bibitem{melville2005sri}
D.~O. Melville and R.~J. Blaikie, \enquote{Super-resolution imaging through a
  planar silver layer,} Opt. Express \textbf{13}, 2127--2134 (2005).

\bibitem{fang2005sdl}
N.~Fang, H.~Lee, C.~Sun, and X.~Zhang, \enquote{Sub-diffraction-limited optical
  imaging with a silver superlens,} Science \textbf{308}, 534--537 (2005).

\bibitem{notomi2000tlp}
M.~Notomi, \enquote{Theory of light propagation in strongly modulated photonic
  crystals: Refractionlike behavior in the vicinity of the photonic band gap,}
  Phys. Rev. B \textbf{62}, 10696--10705 (2000).

\bibitem{zhang2006lst}
H.~Zhang, L.~Shen, L.~Ran, Y.~Yuan, and J.~Kong, \enquote{Layered superlensing
  in two-dimensional photonic crystals,} Opt. Express \textbf{14}, 11178--11183
  (2006).

\bibitem{luo2003sip}
C.~Luo, S.~G. Johnson, J.~D. Joannopoulos, and J.~B. Pendry,
  \enquote{Subwavelength imaging in photonic crystals,} Phys. Rev. B
  \textbf{68}, 45115 (2003).

\bibitem{grbic2004odl}
A.~Grbic and G.~V. Eleftheriades, \enquote{Overcoming the diffraction limit
  with a planar left-handed transmission-line lens,} Phys. Rev. Lett.
  \textbf{92}, 117403 (2004).

\bibitem{zhang2007can}
H.~Zhang, H.~Zhu, L.~Qian, and D.~Fan, \enquote{Collimations and negative
  refractions by slabs of 2d photonic crystals with periodically-aligned
  tube-type air holes,} Opt. Express \textbf{15}, 3519--3530 (2007).

\bibitem{webb2004mnr}
K.~J. Webb, M.~Yang, D.~W. Ward, and K.~A. Nelson, \enquote{Metrics for
  negative-refractive-index materials,} Phys. Rev. E \textbf{70}, 35602 (2004).

\bibitem{smith2002lsd}
D.~R. Smith, D.~Schurig, M.~Rosenbluth, S.~Schultz, S.~A. Ramakrishna, and
  J.~B. Pendry, \enquote{Limitations on subdiffraction imaging with a negative
  refractive index slab,} Appl. Phys. Lett. \textbf{82}, 1506 (2003).

\bibitem{wood2006}
B.~Wood, J.~B. Pendry, and D.~P. Tsai, \enquote{Directed subwavelength imaging
  using a layered metal-dielectric system,} Phys. Rev. B \textbf{74}, 115116
  (2006).

\bibitem{anantharamakrishna2003raa}
S.~A. Ramakrishna and J.~B. Pendry, \enquote{Removal of absorption and increase
  in resolution in a near-field lens via optical gain,} Phys. Rev. B
  \textbf{67}, 201101 (2003).

\bibitem{ponizovskaya2007mni}
E.~V. Ponizovskaya and A.~M. Bratkovsky, \enquote{Metallic negative index
  nanostructures at optical frequencies: losses and effect of gain medium,}
  Appl. Phys. A \textbf{87}, 161--165 (2007).

\bibitem{Fadeyeva:oe2010}
T.~A. Fadeyeva, V.~G. Shvedov, Y.~V. Izdebskaya, A.~V. Volyar, E.~Brasselet,
  D.~N. Neshev, A.~S. Desyatnikov, W.~Krolikowski, and Y.~S. Kivshar,
  \enquote{Spatially engineered polarization states and optical vortices in
  uniaxial crystals,} Opt. Express \textbf{18}, 10848--10863 (2010).

\bibitem{Wang:oe2010}
X.~L. Wang, Y.~Li, J.~Chen, C.~S. Guo, J.~Ding, and H.~T. Wang, \enquote{A new
  type of vector fields with hybrid states of polarization,} Opt. Express
  \textbf{18}, 10786--10795 (2010).

\bibitem{Zhan:aop2009}
Q.~Zhan, \enquote{Cylindrical vector beams: from mathematical concepts to
  applications,} Adv. in Opt. Photon. \textbf{1}, 1--57 (2009).

\bibitem{Ramakrishna:JMO-49-1747}
S.~A. Ramakrishna, J.~B. Pendry, D.~Schurig, D.~R. Smith, and S.~Schultz,
  \enquote{The asymmetric lossy near-perfect lens,} J. Mod. Optics \textbf{49},
  1747--1762 (2002).

\bibitem{Saleh}
B.~Saleh and M.~Teich, \emph{Fundamentals of Photonics} (John Wiley \& Sons,
  Inc, 2007), 2nd ed.

\bibitem{GoodmanFourierOptics}
J.~W. Goodman, \emph{Introduction to Fourier Optics} (Roberts \& Co Publ.,
  2005), 3rd ed.

\bibitem{Moore:josaa2008}
C.~P. Moore, M.~D. Arnold, P.~J. Bones, and R.~J. Blaikie, \enquote{Image
  fidelity for single-layer and multi-layer silver superlenses,} J. Opt. Soc.
  Am. A \textbf{25}, 911--918 (2008).

\bibitem{Scalora:OE09}
N.~Mattiucci, D’Aguanno, M.~Scalora, M.~J. Bloemer, and C.~Sibilia,
  \enquote{Transmission function properties for multi-layered structures:
  Application to super-resolution,} Opt. Express \textbf{17}, 17517--17529
  (2009).

\bibitem{Kotynski:ol2010}
R.~Kotynski and T.~Stefaniuk, \enquote{Multiscale analysis of subwavelength
  imaging with metal-dielectric multilayers,} Opt. Lett. \textbf{35},
  1133--1135 (2010).

\bibitem{Kotynski:jopa2009}
R.~Kotynski and T.~Stefaniuk, \enquote{Comparison of imaging with
  sub-wavelength resolution in the canalization and resonant tunnelling
  regimes,} J. Opt. A: Pure Appl. Opt. \textbf{11}, 015001 (2009).

\bibitem{Kotynski:oer2010}
R.~Kotynski, \enquote{Fourier optics approach to imaging with sub-wavelength
  resolution through metal-dielectric multilayers,} Opto-Electron. Rev.
  \textbf{18}, 366--375 (2010).

\bibitem{Lee-Lalanne-Fainman:ao2010}
B.~Lee, P.~Lalanne, and Y.~Fainman, \enquote{Plasmonic diffractive optics and
  imaging: feature introduction,} Appl. Opt. \textbf{49}, PD01 (2010).

\bibitem{Norfolk:oe2010}
A.~W. Norfolk and E.~J. Grace, \enquote{Reconstruction of optical fields with
  the quasi-discrete {H}ankel transform,} Opt. Express \textbf{18},
  10551--10556 (2010).

\bibitem{Palik}
E.~Palik, ed., \emph{Handbook of Optical Constants of Solids} (Academic Press,
  1998).

\bibitem{JohnsonChristy}
P.~Johnson and R.~Christy, \enquote{Optical constants of the noble metals,}
  Phys. Rev. B \textbf{6}, 4370--4379 (1972).

\end{thebibliography}
\end{document}